\documentclass[aps,prl,twocolumn,superscriptaddress,floatfix]{revtex4}
\usepackage[english]{babel} 
\usepackage{graphicx}
\usepackage{amssymb}
\usepackage{dcolumn}
\usepackage{bm}
\usepackage{amsmath}

\begin{document}

\title{Discrete Nonlinear Breathing Modes in Carbon Nanotubes}

\author{Alexander V. Savin}

\affiliation{Semenov Institute of Chemical Physics, Russian Academy
of Sciences, Moscow 119991, Russia}

\author{Yuri S. Kivshar}

\affiliation{Nonlinear Physics Center, Research School of Physical
Sciences and Engineering, Australian National University, Canberra,
ACT 0200, Australia}

\begin{abstract}
We study large-amplitude oscillations of carbon nanotubes with
chiralities $(m,0)$ and $(m,m)$ and predict the existence of
localized nonlinear modes in the form of {\em discrete breathers}.
In nanotubes with the index $(m,0)$ {\em three types} of localized
modes can exist, namely longitudinal, radial, and twisting
breathers; however only the twisting breathers, or {\em twistons},
are nonradiating nonlinear modes which exist in the frequency gaps
of the linear spectrum. Geometry of carbon nanotubes with the index
$(m,m)$ allows only the existence of broad radial breathers in a
narrow spectral range.
\end{abstract}


\maketitle

Carbon nanotubes~\cite{book} are attracting considerable attention
in recent years after their discovery by Iijima~\cite{disc}. They
can be thought of as cylinders of carbon atoms arranged in hexagonal
grids and are thus fullerene-related structures. The growing
interest to study carbon nanotubes can be explained by their unique
physical properties and their potential for a wide range of possible
applications. In particular, the carbon nanotubes are known for
their superior mechanical strength~\cite{ref_1} and good heat
conductance~\cite{ref_2}. In addition, it is well established that
C$_{60}$ fullerenes can support large-amplitude
oscillations~\cite{rao} which can be excited and controlled by
temporally shaped laser pulses~\cite{laa_prl}.

In the continuum approximation, nonlinear dynamics of carbon
nanotubes has been analyzed by several groups and, in particular,
supersonic longitudinal compression solitons described by the
effective Korteweg-de Vries equation have been predicted to exist in
such structures, similar to other simpler discrete
lattices~\cite{ast_prb}. However, the recent numerical modeling of
more complete discrete model of carbon nanotubes has
demonstrated~\cite{savin} that acoustic solitons {\em do not exist}
in such curved structures, and their supersonic motion is always
accompanied by strong radiation of phonons.

In this Letter, we focus on the study of {\em large-amplitude}
oscillating modes of carbon nanotubes that have the additional
features of being {\em nonlinear} as well as {\em discrete}. We
reveal that both nonlinearity and discreteness induce localization
of anharmonic oscillations and, as a result, the combination of both
leads to the generation of specific spatially localized modes of the
type of discrete breathers~\cite{sie_tak,mac_aub}. These modes act
like stable effective impurity modes that are dynamically generated
and may alter dramatically many properties of carbon nanotubes.

Discrete breathers appear in strongly nonlinear systems, and their
spatial size become comparable to the lattice spacing. Such discrete
breathers--also called `intrinsic localized modes' or `discrete
solitons' are responsible for energy localization in the dynamics of
discrete nonlinear lattices~\cite{flach,phys_today}. The
manipulation of the discrete breathers has been achieved in systems
as diverse as annular arrays of coupled Josephson
junctions~\cite{jj}, optical waveguide arrays~\cite{optics}, and
antiferromagnetic spin lattices~\cite{sievers}. The direct
observation of highly localized, stable, nonlinear excitations at
the atomic level in the structures such as carbon nanotubes will
underscore their importance in physical phenomena at all scales. In
this Letter we demonstrate that carbon nanotubes with the index
$(m,0)$ support at least {\em three types} of discrete breathers,
whereas nanotubes with the index $(m,0)$ support only one type of
discrete breather.

%---------------------------- Fig. 1 ------------------------------------
\begin{figure}[b]
\includegraphics[angle=0, width=.8\linewidth]{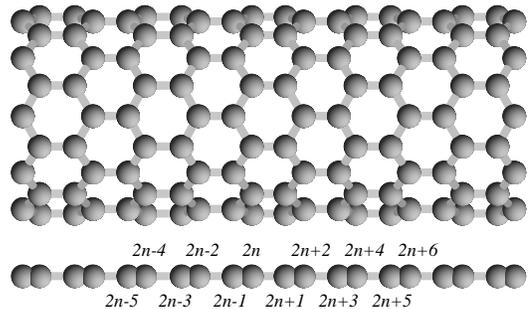}
%\centerline{
%    \scalebox{0.5}{
%      \includegraphics{fig1.eps}}}
\caption{Schematic of a carbon nanotube with the index $(m,0)$.
Below: an effective one-dimensional diatomic chain. }
\label{fig1}
\end{figure}
%---------------------------- Fig. 1 ------------------------------------
The structure of a carbon nanotube is shown schematically in Fig.~1.
In static, the nanotube is characterized by its radius $R$ and two
step parameters $h_1$ and $h_2$. In each layer, the nanotube has $m$
atoms separated by the angular distance $\Delta \phi = 2\pi/m$, so
that $h_1$ and $h_2$ define alternating longitudinal distances
between the transverse layers. We consider such dynamics of the
nanotube that all atoms in one transverse layer have identical
displacements. In this case, the carbon nanotube can be modeled by
an effective one-dimensional diatomic chain, where the coordinates
of atoms
$(\rho_{n,l}\cos\varphi_{n,l},~\rho_{n,l}\sin\varphi_{n,l},~z_n)$
are defined by the equations: $\rho_{n,l}=R+r_n(t)$;
$\varphi_{n,l}=\Delta\phi(l-1)+\phi_n(t)$,
$z_{n,l}=(2k-2)(h_1+h_2)+u_n(t)$ for $n=4k-1$;
$\varphi_{n,l}=\Delta\phi(l-1)+\Delta\phi/2+\phi_{n}(t)$,
$z_{n,l}=(2k-2)(h_1+h_2)+h_1+u_n(t)$ for $n=4k$;
$\varphi_{n,l}=\Delta\phi(l-1)+\Delta\phi/2+\phi_{n}(t)$,
$z_{n,l}=(2k-1)(h_1+h_2)+u_n(t)$ for $n=4k+1$; and
$\varphi_{n,l}=\Delta\phi(l-1)+\phi_{n}(t)$,
$z_{n,l}=(2k-1)(h_1+h_2)+h_1+u_n(t)$ for $n=4k+2$, where the index
$n=4k+i$, $(k=0,\pm 1,\pm2,...,~i=-1,0,1,2)$ stands for the number
of the transverse atomic layer, the index $l=1,...,m$ marks an atom
in the transverse layer, $r_n(t)$ is a relative change of the radius
of the $n$-th transverse layer, $\phi_n$ is the angle of rotation of
the atoms in the layer, and $u_n$ is a relative longitudinal
displacement of the atoms from their equilibrium position. In the
static case, $r_n\equiv 0$, $\phi_n\equiv 0$, and $u_n\equiv 0$.

In this case, complex three-dimensional dynamics of a nanotube can
be reduced to the analysis of an effective one-dimensional model
with two atoms per its unit cell, and Hamiltonian of this model can
be written in the form,
\begin{equation}
H{=}\sum_n[E_{2n}{+}E_{2n+1}{+}Z({\bf y}_{2n-2};{\bf y}_{2n-1};
{\bf y}_{2n};{\bf y}_{2n+1})],
\label{f1}
\end{equation}
where
$E_n=M[\dot{r}_{n}^2+(R+r_{n})^2\dot{\phi}_{n}^2+\dot{u}_{n}^2]/2$
is the kinetic energy , $M$ is the mass of a carbon atom (the total
energy of the nanotube is $mH$), and the vector ${\bf
y}_n=(r_n,\phi_n,u_n)$. The interatomic potential can be written in
the form,
\begin{eqnarray}
Z({\bf y}_1,{\bf y}_2,{\bf y}_3,{\bf y}_4)=
V({\bf x}_1,{\bf x}_2)+V({\bf x}_2,{\bf x}_3)+V({\bf x}_2,{\bf x}_4)
~\nonumber\\
+U({\bf x}_2,{\bf x}_3,{\bf x}_5)+U({\bf x}_6,{\bf x}_3,{\bf x}_2)
+U({\bf x}_6,{\bf x}_3,{\bf x}_5)~\nonumber\\
+U({\bf x}_1,{\bf x}_2,{\bf x}_3)
+U({\bf x}_1,{\bf x}_2,{\bf x}_4)+U({\bf x}_3,{\bf x}_2,{\bf x}_4)~\nonumber\\
{+}W({\bf x}_6,{\bf x}_3,{\bf x}_2,{\bf x}_5)
{+}W({\bf x}_6,{\bf x}_3,{\bf x}_5,{\bf x}_2)
{+}W({\bf x}_2,{\bf x}_3,{\bf x}_6,{\bf x}_5)~\nonumber\\
{+}W({\bf x}_1,{\bf x}_2,{\bf x}_4,{\bf x}_3)
{+}W({\bf x}_1,{\bf x}_2,{\bf x}_3,{\bf x}_4)
{+}W({\bf x}_3,{\bf x}_2,{\bf x}_1,{\bf x}_4),\nonumber
\end{eqnarray}
where the coordinate vectors ${\bf x}_i=(x_i,y_i,z_i)$,
$i=1,2,...,6$ are defined as: $x_6=(R+r_1)\cos(\Delta\phi+\phi_1)$,
$y_6=(R+r_1)\sin(\Delta\phi+\phi_1)$, $z_6=-h_2+u_1$,
$x_1=(R+r_2)\cos(\phi_2)$, $y_1=(R+r_2)\sin(\phi_2)$, $z_1=u_2$,
$x_3=(R+r_2)\cos(\Delta\phi+\phi_2)$,
$y_3=(R+r_2)\sin(\Delta\phi+\phi_2)$, $z_3=u_2$,
$x_2=(R+r_3)\cos(\Delta\phi/2+\phi_3)$,
$y_2=(R+r_3)\sin(\Delta\phi/2+\phi_3)$, $z_2=h_1+u_3$,
$x_5=(R+r_3)\cos(3\Delta\phi/2+\phi_3)$,
$y_5=(R+r_3)\sin(3\Delta\phi/2+\phi_3)$, $z_5=h_1+u_3$,
$x_4=(R+r_4)\cos(\Delta\phi/2+\phi_4)$,
$y_4=(R+r_4)\sin(\Delta\phi/2+\phi_4)$, $z_4=h_1+h_2+u_4$.
The potential
$V({\bf x}_1,{\bf x}_2)=D\{\exp(-\alpha[\rho-\rho_0])-1\}^2$,~
$\rho=|{\bf x}_2-{\bf x}_1|$,
%\label{f3}
%\end{equation}
%
describes a change of the deformation energy due to interaction
between two atoms with the coordinates ${\bf x}_1$ and ${\bf x}_2$.
Potential $U({\bf x}_1,{\bf x}_2,{\bf
x}_3)=\epsilon_v(\cos\varphi+1/2)^2$, where $\cos\varphi=({\bf
v}_1,{\bf v}_2)/(|\bf{v}_1|\cdot|\bf{v}_2|)$, and ${\bf v}_1={\bf
x}_2-{\bf x}_1$, ${\bf v}_2={\bf x}_3-{\bf x}_2$, describes the
deformation energy of the angle between the links ${\bf x}_1{\bf
x}_2$ and ${\bf x}_2{\bf x}_3$. Finally, the potential $W({\bf
x}_1,{\bf x}_2,{\bf x}_3,{\bf x}_4)=\epsilon_t(1-\cos\phi)$, where
$\cos\phi=(\bf{u}_1,\bf{u}_2)/(|\bf{u}_1|\cdot|\bf{u}_2|)$ and
$\bf{u}_1=({\bf x}_2-{\bf x}_1)\times({\bf x}_3-{\bf x}_2)$,
$\bf{u}_2=({\bf x}_3-{\bf x}_2)\times({\bf x}_4-{\bf x}_3)$,
describes the deformation energy associated with a change of the
effective angle between the planes ${\bf x}_1{\bf x}_2{\bf x}_3$ and
${\bf x}_2{\bf x}_3{\bf x}_4$. We take the mass of carbon atom as
$M=12m_p$, where $m_p$ if the proton mass, the length
$\rho_0=1.418$\AA, and energy $D=4.9632$~eV. Other model parameters
such as $\alpha$, $\epsilon_v$, and $\epsilon_t$ can be determined
from the phonon frequency spectrum of a plane of carbon atoms.

A flat plane of the carbon atoms (graphene) is a special case of
carbon nanotubes in the limit $R\rightarrow\infty$, when
$h_1=\rho_0/2$ and $h_2=\rho_0$. For such a plane the motion
equation splits into the equations for longitudinal and transverse
motion. The corresponding Hamiltonian takes a simpler form,
\begin{equation}
H=\sum_n[\frac12M(\dot{u}_{2n-1}^2+\dot{u}_{2n}^2)+V_1(\rho_{2n})
+V_2(\rho_{2n-1})],
\label{f2}
\end{equation}
where $\rho_n=u_{n+1}-u_n$, the potentials $V_1(w)=D[\exp(-\alpha w)-1]^2$,
$V_2(w)=2D\{\exp(-\alpha [a_+(w)^{1/2}-\rho_0])-1\}^2
+2\epsilon_v[a_-(w)/a_+(w)+1/2]^2+4\epsilon_v[(w+\rho_0/2)
/\sqrt{a_+(w)}-1/2]^2$,
and function $a_\pm(w)=(w+\rho_0/2)^2\pm 3\rho_0^2/4$.

\begin{figure}[tbh]
\begin{center}
\includegraphics[angle=0, width=0.95\linewidth]{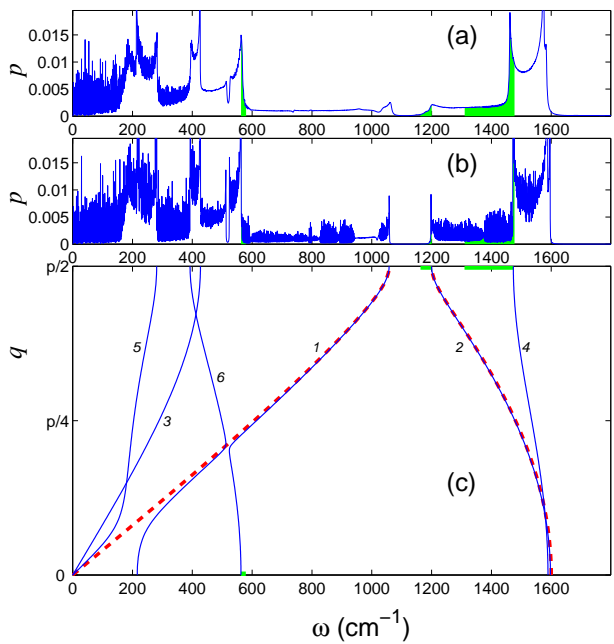}
\end{center}
\caption{\label{fig2} Spectral density of thermal oscillations of a
carbon nanotube $(10,0)$ for the temperature (a) $T=300$K and (b)
$T=30$K. (c) Dispersion curves of the phonons: acoustic (1) and
optical (2) longitudinal phonons, acoustic (3) and optical (4)
rotation phonons, and acoustic (5) and optical (6) radial phonons.
For comparison, red dashed lines show the dispersion curves of
longitudinal oscillations of a plane of carbon. Green color marks
the frequency spectrum of breathers.}
\end{figure}

The stiffness parameters of the potentials are
$K_1=V_1''(0)=2D\alpha^2$ and
$K_2=V_2''(0)=D\alpha^2+27\epsilon_v/2\rho_0^2$. After linearizing
the equation of motion following from the Hamiltonian (\ref{f2}), we
obtain the dispersion relation for longitudinal phonons,
$\omega_\pm(q)=\{(K_1+K_2\pm[(K_1+K_2)^2-2K_1K_2(1-\cos
2q)]^{1/2})/M\}^{1/2}$, which are depicted in Fig.~\ref{fig2}. The
frequency spectrum consists of the acoustic
$[\omega_-(0),\omega_-(\pi/2)]$ and optical
$[\omega_+(\pi/2),\omega_+(0)]$ bands where $\omega_-(0)=0$,
~~$\omega_+(0)=\sqrt{2(K_1+K_2)/M}$,
~~$\omega_-(\pi/2)=\sqrt{2K_2/M}$, and
~~$\omega_+(\pi/2)=\sqrt{2K_1/M}$.

The edge frequencies of the optical band can be estimated from the
experimental data: $\omega_+(\pi/2)\approx 1200$~cm$^{-1}$ and
$\omega_+(0)\approx 1600$~cm$^{-1}$~\cite{p1}. These values allow us
to determine the stiffness parameters, $K_1=508.98$~N/m and
$K_2=395.87$~N/m, and find the maximum frequency of acoustic
phonons, $\omega_-(\pi/2)=1058.3$~cm$^{-1}$. Knowing the values of
$K_1$ and $K_2$, we then find other parameters,
$\alpha=1.7889$~\AA$^{-1}$ and $\epsilon_v=1.3143$~eV. The value of
the torsion potential $\epsilon_t$ can be evaluated from the maximum
frequency of the transverse oscillations of a plane carbon lattice.
For $\epsilon_t=0.2$~eV, we find the value 570 cm$^{-1}$. In order
to find the parameters $R$, $h_1$, and $h_2$, we should solve the
minimum problem $Z({\bf 0},{\bf 0},{\bf 0},{\bf
0})\rightarrow\min_{R,h_1,h_2}$. The resulting value of energy is
then used as the minimum value.

\begin{figure}[tbh]
\begin{center}
\includegraphics[angle=0, width=1.0\linewidth]{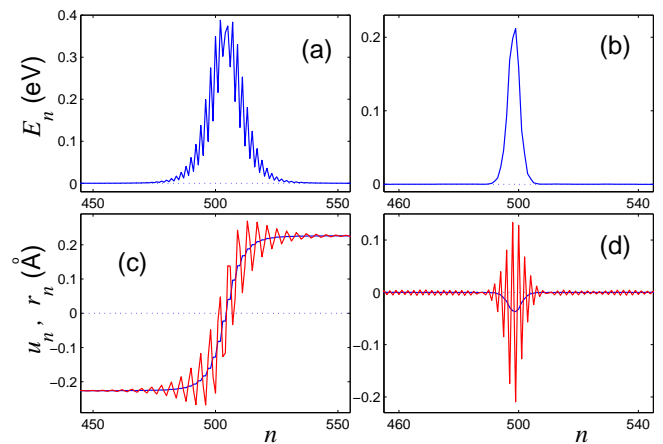}
\end{center}
\caption{\label{fig3} Example of a localized nonlinear mode of the
longitudinal oscillations described by the Hamiltonian (\ref{f2}) (a), (b)
(frequency $\omega=1164$cm$^{-1}$, energy $E=5.71$eV, width
$L=16.3$, the chain extension $A=0.45$\AA)
and example of a radial breather (c), (d) describing localized transverse 
oscillations of a nanotube (10,0) with the frequency $\omega=579.6$~cm$^{-1}$, 
energy $E=0.99853$eV, and width $L=32.7$.
Shown are (a) and (c) the averaged (in time) energy distribution $E_n$ in
the chain, (b) the atom displacements $u_n$, and (d) transverse displacements 
$r_n$. In sections (b) and (d) blue lines show the values averaged over the 
period, red -- maximal displacements.
Radiation of longitudinal waves by a radial breather
is clearly visible in section (d).
}
\end{figure}

A simple form of the Hamiltonian (\ref{f2}) allows to obtain
analytical results for the nonlinear dynamics similar to the case of
diatomic lattices. These results allow to predict the existence of
discrete breathers with the frequencies below the lowest optical
frequency of the longitudinal phonons [see Fig.~\ref{fig2}(c)]. The
form of this breather is shown in Fig.~\ref{fig3}. The breather is
characterized by the frequency $\omega$, energy $E$, width of the
localization region, $L$, and the chain extension $A$. The breather
frequency is inside the band [1162, 1200]~cm$^{-1}$ near the lowest
edge of the longitudinal optical oscillations. For decreasing
$\omega$, both $E$ and $A$ grow monotonically, and the breather
width decreases.
%---------------------------- Fig. 2 ------------------------------------
%---------------------------- Fig. 2 ------------------------------------

Hamiltonian (\ref{f1}) defines the motion equations of the system.
We have studied these equations numerically and revealed that they
support the existence of {\em three types} of strongly localized
nonlinear modes--{\em discrete breathers}. The first type, {\em
longitudinal breathers}, also exists in planar carbon structures,
such breathers exist in the frequency range [1162, 1200]~cm$^{-1}$.
The second type, {\em radial breathers}, describes transverse
localized nonlinear modes with the frequency band
[562, 580]~cm$^{-1}$. The third type, {\em twisting breathers},
characterizes localization of the torsion oscillations of the
nanotube with the frequencies [1310, 1477]~cm$^{-1}$.  The frequency
spectra of the breathers are shown in Fig.~\ref{fig2}.

For a flat plane of carbon atoms, the longitudinal breathers are
nonlinear modes [see Figs.~\ref{fig3}(a,c)], and they are exact
solutions of the nonlinear motion equations. However, in the case of
a curved geometry,  the longitudinal breathers become coupled to
transverse linear modes, and they always emit some radiation. This
radiation is defined by the curvature of the nanotube and its index
$m$. Therefore,  the longitudinal breathers are not genuine
nonlinear modes of carbon nanotubes, and they possess a finite
lifetime which however may exceed a few hundred of picoseconds.

%---------------------------- Fig. 3 ------------------------------------
%---------------------------- Fig. 3 ------------------------------------
The second type of discrete breathers we found is associated with
the localization of transverse radial oscillations of a nanotube.
Example of this radial breather in the nanotube (10,0) is shown in
Figs.~\ref{fig3}(b,d). Localized out-phase transverse oscillations
of the neighboring atoms lead to localized contraction and extension
of the nanotube. Such transverse oscillations become coupled to the
longitudinal oscillations and, therefore, the radial breathers radiate
longitudinal phonons. As a result, the radial breathers are also not
genuine nonlinear localized modes of the carbon nanotubes, and they
decay slow by emitting small-amplitude phonons. The lifetime of
these breathers can be of the order of several nanoseconds.

\begin{figure}[th]
\begin{center}
\includegraphics[angle=0, width=.8\linewidth]{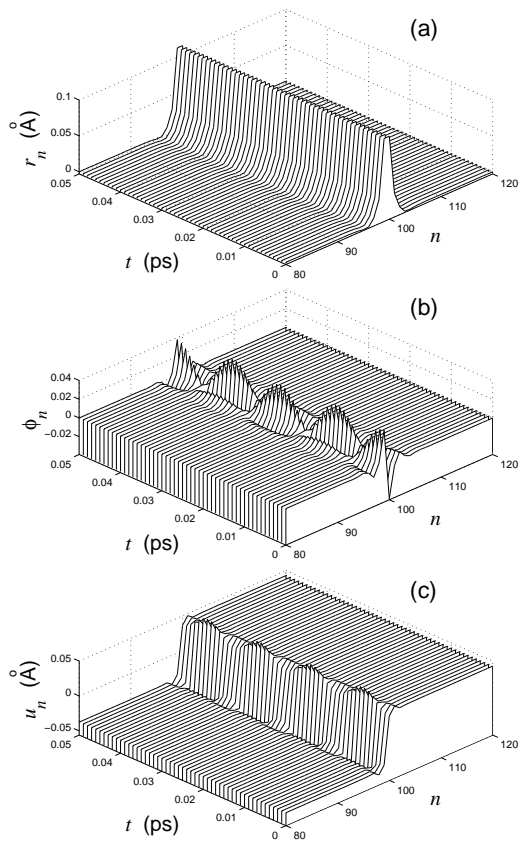}
\end{center}
\caption{\label{fig4} Dependence of (a) transverse $r_n$, (b) torsion $\phi_n$,
and (c) longitudinal $u_n$ displacements of the twisting discrete breather
(frequency $\omega=1367$~cm$^{-1}$, energy $E=1.6465$~eV, width
$L=1.55$) on the time $t$.
}
\end{figure}

\begin{figure}[tbh]
\begin{center}
\includegraphics[angle=0, width=1.\linewidth]{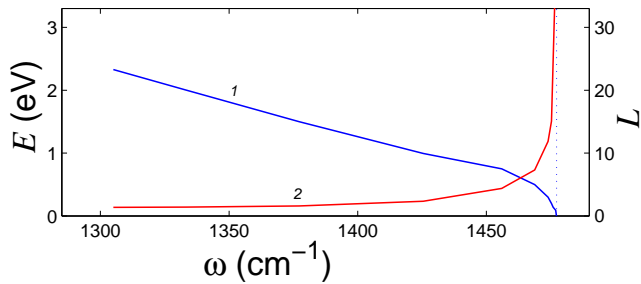}
\end{center}
\caption{\label{fig5} Dependence of total energy $E$ (curve 1) and width $L$
(curve 2) of the twisting discrete breather on the frequency $\omega$. 
Vertical line show the edge of the linear spectrum of optical torsion 
oscillations in the
(10,0) nanotube.}
\end{figure}

The third type of localized mode is a twisting breather, or {\em
twiston}, associated with the torsion oscillations of the nanotube.
In a sharp contrast to other two breathing modes, the twisting
breather is an exact solution of the motion equations of the
nanotube, and it does not radiate phonons. An example of this
genuine discrete breather is shown in Fig.~\ref{fig4}. In the
localized region of this mode, the nanotube is expanded
transversally being contracted longitudinally. The twiston has a
broad frequency spectrum, and its energy, amplitude of the
transverse extension (see Fig.~\ref{fig5}), and the amplitude of
torsion oscillations all grow with the frequency. The breather width
changes monotonically, and for the frequencies $\omega<1450$
cm$^{-1}$ it becomes comparable with the lattice spacing, so that
the breather becomes a highly localized mode.
%---------------------------- Fig. 4 ------------------------------------
%---------------------------- Fig. 4 ------------------------------------
%---------------------------- Fig. 5 ------------------------------------
%---------------------------- Fig. 5 ------------------------------------

Next, we analyze thermal oscillations of a carbon nanotube by
employing the Langevin equations. We find that for low temperatures
($T=30$K) the oscillations are mostly linear and  the frequency
density does not differ much from the density of linear modes, as
demonstrated in Fig.~\ref{fig2}(b). However, for higher temperatures
($T=300$K) the spectral density acquires specific features
associated with the generation of nonlinear modes, as can be seen in
Fig.~\ref{fig2}(a) where we observe thermal oscillations with the
frequencies inside the linear spectral gaps which can be associated
with discrete breathers. Indeed, the larger contribution of these
nonlinear modes is for the torsion oscillations of the twisting
breathers, which are stable and have the largest frequency spectrum.

We have carried out the similar nonlinear analysis for the nanotubes
with the index $(m,m)$ and revealed that in this nanotube can
support only one type of breathers, a radial breather with a very
narrow frequency spectrum [430.5, 436]~cm$^{-1}$ near the upper edge
of the frequency band of radial phonons. However, this radial
breather is not an exact solution of the nonlinear motion equations,
and radiation of small-amplitude linear waves leads to a decay of
the breather. As a result, the existence of nonlinear localized
modes depends crucially on chirality of the carbon nanotube, so that
genuine discrete breathers are expected to exist in the nanotube
with the index $(m,0)$. Existence of twisting breathers is due to
the anharmonic interaction potential, and the large spectral gap in
the frequency spectrum of torsion phonons. However, the radial
long-lived nonlinear modes can appear in the nanotubes with any type
of chirality.

In conclusion, we have revealed that carbon nanotubes can support
spatially localized large-amplitude stable nonlinear modes in the
form of discrete breathers, and we have analyzed the existence and
stability of three types of breathers. A novel type of such highly
localized discrete modes --twistons-- is associated with the energy
self-trapping of torsion oscillations of the carbon nanotubes.

This work was supported by the Australian Research Council. Alex
Savin thanks the Nonlinear Physics Center of the Australian National
University for a warm hospitality during his stay in Canberra.

\end{document}